\documentclass{acmart}

\usepackage[english]{babel}
\usepackage[utf8x]{inputenc}
\usepackage{amsmath}
\usepackage{graphicx}
\usepackage[colorinlistoftodos]{todonotes}
\usepackage{url}
\usepackage{subfigure}
\usepackage{multirow}
\usepackage{amssymb}

\begin{document}
\title{On labeling Android malware signatures using minhashing and further classification with Structural Equation Models}


\author{Ignacio Mart\'{i}n}
\affiliation{
  \institution{Universidad Carlos III de Madrid}
  \streetaddress{Avda de la Universidad 30}
  \city{Leganés} 
  \state{Madrid}
  \country{Spain}
  \postcode{28911}
}
\email{ignmarti@it.uc3m.es}

\author{Jos\'{e} Alberto Hern\'{a}ndez}
\affiliation{
  \institution{Universidad Carlos III de Madrid}
  \streetaddress{Avda de la Universidad 30}
  \city{Leganés} 
  \state{Madrid}
  \country{Spain}
  \postcode{28911}
 }
\email{jahgutie@it.uc3m.es}

\author{Sergio de los Santos}
\affiliation{%
  \institution{Telef\'{o}nica Digital Identity \& Privacy}
  \streetaddress{Ronda de la Comunicación}
  \city{Madrid} 
  \country{Spain}
}
\email{ssantos@11paths.com}

\begin{abstract}
Multi-scanner Antivirus systems provide insightful information on the nature of a suspect application; however there is often a lack of consensus and consistency between different Anti-Virus engines. In this article, we analyze more than 250 thousand malware signatures generated by 61 different Anti-Virus engines after analyzing 82 thousand different Android malware applications. We identify 41 different malware classes grouped into three major categories, namely Adware, Harmful Threats and Unknown or Generic signatures. We further investigate the relationships between such 41 classes using community detection algorithms from graph theory to identify similarities between them; and we finally propose a Structure Equation Model to identify which Anti-Virus engines are more powerful at detecting each macro-category. As an application, we show how such models can help in identifying whether Unknown malware applications are more likely to be of Harmful or Adware type.
\end{abstract}


\maketitle

\section{Introduction}
\label{sec:intro}
Smartphones and tablets have become part of our daily life. The number of such smart devices keeps growing year after year\footnote{See \url{http://www.smartinsights.com/mobile-marketing/mobile-marketing-analytics/mobile-marketing-statistics/}, last access: March 2017}. Android is the most popular mobile operating system and has grown into a diverse ecosystem worldwide. 

Unfortunately, the success of Android has also attracted malware developers: it is estimated that about a 12\% of apps in the Google play market are "low quality apps"\footnote{See \url{http://www.appbrain.com/stats/number-of-android-apps}, last access: July 2017}, many of them represent a real risk for the smartphone owner. 


There exist in the literature many systems to detect Android malware using classical detection approaches. For example, in \cite{ANDRUBIS}, the authors have developed an Android Malware analysis tool and review malware behavior based on a 1-Million sample of Android applications, highlighting differences between malware and goodware. Elish et al~\cite{profilingUsers} have proposed a single-feature classification system based on user behavior profiling. In general, Android permissions have had a wide coverage and works like \cite{puma,permissionsDemystified} analyze them in detail.

Concerning the use of Machine Learning (ML) techniques in the detection of malware, the authors in \cite{drebin} have gathered features from application code and manifest (permissions, API calls, etc) and use Support Vector Machines (SVMs) to identify different types of malware families. In a different approach, the authors of \cite{massVetting} have proposed a system based on the differential-intersection analysis of applications in order to spot duplicates. 

Antivirus Software has been persistently analyzed and tested. For instance, the authors in \cite{AVMobileDesign} have reviewed the key points in designing AV engines for mobile devices as well as how to avoid detection. In a different approach, Rastogi et al~\cite{catchme} have assessed whether AV engines fall for obfuscation attacks, finding many to be vulnerable to some kind of transformation attack. The authors in~\cite{CCSPoster} have performed data analytics on multi-scanner outputs for Android Applications to find their behavior patterns. 


With the advent of AV multi-scanner tools, such as Meta-Scan, Virustotal or Androguard, any application can easily be analyzed by many different AV engines at once. For each detected application, these tools typically identify the AV engines who flagged the application as malware, its type and other meta-data regarding the nature of the threat. Hence, multi-scanner tools enable simultaneous analysis of suspicious applications and provides some information to identify and deal with many types of malware.

The authors in~\cite{AVComparisonVT} perform a comparison of AV engines from VirusTotal by modeling AV confidence using a hyper-exponential curve. In~\cite{StudyAVs}, AV labels from VirusTotal are subject to temporal analysis using a collection of malware applications obtained through a honeypot network. Additionally, other studies \cite{malwareDetImprovement,diverseProtection} have shown the advantages of using more than one AV engine to improve malware decisions, by means, for example, of multi-scanner tools.

Nevertheless, the authors in \cite{lackConsensus} recall the lack of agreement on which application each AV considers as malware. Besides, Maggi et al.~\cite{nameInconsistencies} extensively review the inconsistencies when assigning identifiers to similar threats across engines. In this light, the authors in \cite{betterMalwareGT} propose a combination scheme for multi-scanner detections based on a Generative Bayesian model to infer the probability of being malware for every sample, however no specific label analysis is performed, and thus all threats are treated equally. 


Several authors have analyzed and proposed categorization schemes for Android malware applications. In \cite{malwareFamilies} the authors find up to 49 distinct malware families whilst the authors in \cite{dendroid} propose a text mining approach to obtain and classify malware families according to application code. Similarly, Zheng et al propose in \cite{droidAnalytics} a system for the collection and categorization of zero-day malware samples into different families. Also, the authors in~\cite{droidLegacy} propose a system to classify malware samples according to their families.


Sebastián et al.~\cite{avclass} propose AVClass, a system to normalize AV labels from different vendors and determine the actual class out different detection outputs for the same applications. Nevertheless, AVClass does not link AV engines with their detections. Instead, it provides the frequency for each token and chooses the most probable one. Besides, AVClass removes common malware-related tokens. This way, tokens such as \emph{Adware} or \emph{Trojan} are removed and the information they carry is missed. Consequently, the output of AVClass gives a final malware class output, but loses information on (AV, class) pairs in the process.

In this light, we develop an alternative label normalization methodology based on the well-known minhashing technique~\cite{minhashing}. This system relies on the user to finally assign normalized labels by using python regular expressions over signatures. This way, unsupervised aggregation of signatures can be achieved, considerably reducing the supervising effort of the researcher. 

Then, such methodology will enable cross-engine analysis of malware classes to improve malware classification. In a nutshell, This work contributes to this aim with a twofold effort:
\begin{enumerate}
\item We develop a methodology for signature normalization; that is, group together identifiers referring to the same threat but differing on the actual labels because of AV engine inconsistencies.
\item We model AV engine relationships using Structural Equation Models (SEM) across malware categories aiming at the improvement malware classification.
\end{enumerate}


The rest of this paper is structured as follows: Section~\ref{sec:dataset} describes the data collection and our AV signature normalization methodology. Section~\ref{sec:grouping} inspects engines and signature tokens using correlations to unveil consensual subsets of entities. Section~\ref{sec:logReg} develops different weighting models to evaluate engine performance of distinct malware categories. Finally, Section~\ref{sec:conclusions} summarizes the main findings of this work and highlights the most relevant conclusions.

%
%

\section{Dataset Description and Signature Normalization}
\label{sec:dataset}

In this article, the dataset under study comprises a total of $82,866$ different Android applications collected from Google Play by TACYT\footnote{See \url{https://www.elevenpaths.com/es/tecnologia/tacyt/index.html} for further details} in May 2015. All these applications are considered suspicious, as they have been flagged by at least one out of $61$ antivirus (AV) engines, including some of the most popular ones (e.g. McAffee, Trend Micro, etc.) as well as many others. These engines have been anonymized to preserve privacy, i.e. every engine has been substituted consistently by one of the names in the range $AV_1,\ldots, AV_{61}$ throughout the paper. 

When a malware engine detects a suspicious application, it provides a signature containing some meta-data, like its last scan date or malware class identifier. A total of $259,608$ signatures are obtained in our dataset (i.e. $3.13$ signatures per application on average). 


As an example, consider application no. $1,345$, flagged by AV27, AV28 and AV58. Each AV engine provides different signatures, namely:
\begin{itemize}
\item AV27: \emph{a variant of Android/AdDisplay.Startapp.B}
\item AV28: \emph{Adware/Startapp.A}
\item AV58: \emph{Adware.AndroidOS.Youmi.Startapp (v)} 
\end{itemize}
Clearly, all three engines consider app no. $1,345$ as an adware-like application, but the signature name convention is different for each engine. Thus, text processing and text mining techniques are necessary to convert signatures into a common format for analysis.
 
\subsection{Cleaning and classification of AV signatures with Minhashing}

\begin{figure*}[!htbp]
\centering
\includegraphics[width=0.9\textwidth]{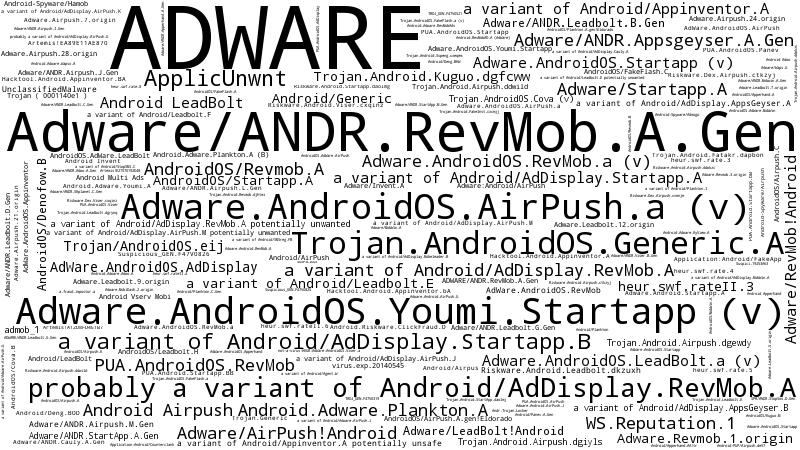}
\caption{Wordcloud image of different raw signatures across the dataset}
\label{fig:signatureCloud}
\end{figure*}

Fig.~\ref{fig:signatureCloud} shows a wordcloud with the most popular AV-generated raw signatures and their frequencies (most popular keywords are shown with large font sizes). Apart from some common understandable signatures, most of them include different names which account for different types of malware and other non-malicious names (i.e. AndroidOS).

Some signatures contain common substrings across different AVs, including related chunks of text regarding very common malware types such as \emph{"PUA"} or \emph{"Trojan"} to more specific types such as \emph{"GingerMaster"} or \emph{"FakeFlash"} together with some related terms which do not refer to malware, namely \emph{AndroidOS} or \emph{win32}. 

To extract meaningful information from signatures, we have developed a methodology to clean, unify and normalize detection identifiers into a fixed subset of \emph{"identifier tokens"} representing the most frequent keywords contained within the signatures. This process starts with conventional text-mining cleaning techniques of raw strings,  including lower-casing, removing punctuation and domain-specific stop-words (i.e. tokens providing no malware information) and splitting each signature into tokens separated by dots. Up to this point, our methodology follows the steps of AVClass~\cite{avclass}.

Next, we use the well-known minhashing algorithm to group signatures together. The \emph{hashing trick} or \emph{minhashing} is a very fast algorithm for estimating how similar (in terms of Jaccard similarity) two sets are. Minhashing relies on splitting strings into several chunks of the same length and computing a unique-output function (i.e. common hash functions like MD5 or SHA1) for each chunk. Consequently, each signature produces a set of numbers, the minimum of which is selected as the minhash. Finally, elements are grouped according to their minhashing values. Once the minhashing values are computed, the probability of two signatures falling in the same group is shown to approximate the Jaccard distance between them. The Jaccard distance between two sets A and B follows:
$$J(A,B) = \frac{|A\cap B|}{|A\cup B|}$$
In other words, similar items will likely fall into similar minhash buckets. A detailed explanation of Minhashing, Jaccard distance and all these terms may be found in~~\cite{minhashing}.

We manually checked the resulting groups and developed a set of Python regular expressions to transform signatures into malware classes according to the unveiled patterns. Since different signatures might contain different classes of malware, collisions may eventually occur within these rules. In this light, we established rule priority following first match criteria over the sorted rules (in terms of specificity). For instance, consider the signature: \emph{"Adware.Android.AirPush.K"}. This signature would fall into the category Airpush, since it is more specific than Adware.

As a result, the generated classes group together similar pattern signatures into a representative set of malware classes. In contrast to AVClass, our approach keeps track of the relationship between the AV and the malware class associated to the signatures.

\subsection{Normalized Signatures}

\begin{table*}[!htpb]
\centering
\begin{tabular}{| c c c c c c c |}
\hline
\# & Regexp rule & Class & Category & Det. Count & No. Apps & AVs \\
\hline
\hline
 S1 & .*a{[}ir{]}*push?.* & Airpush & \multirow{17}{*}{Adware} & 35,850 & 12,802 & 26\\
S2 & .*leadbolt.* & Leadbolt && 17,414 & 4,045 & 21 \\
S3 & .*revmob.* & Revmob && 38,693 & 13,680 & 18 \\
S4 & .*startapp.* & StartApp && 29,443 & 11,963 & 13 \\
S5 & {[}os{]}*apperhand.* \textbar .*counterclank.* & Apperhand/Counterclank && 1,606 & 716 & 12 \\
S6 & .*kuguo.* & Kuguo & & 2,127 & 1,893 & 23\\
S7 & wapsx? & WAPS & &1,546 & 344 & 6\\
S8 & .*dowgin.*\textbar dogwin & Dogwin & & 1,098 & 421 & 23 \\
S9 & .*cauly.* & Cauly &  & 1,143 & 626 & 3 \\
S10 & {[}os{]}*wooboo & Wooboo & & 220 & 120 & 14\\
S11 & {[}os{]}*mobwin & Mobwin & & 1,284 & 249 & 3 \\
S12 & .*droidkungfu.* & DroidKungFu & & 105 & 54 & 3 \\
S13 & .*plankton.* & Plankton & & 4,557 & 741 & 25 \\
S14 & {[}os{]}*you?mi & Youmi & & 1,472 & 370 & 22 \\
S15 & {[}osoneclick{]}*fraud &  Fraud & & 736 & 382 & 19 \\
S16 & multiads & Multiads & & 560 & 555 & 3 \\
S17 & .*adware.*\textbar ad.+ & Adware (gen) & & 33,133 & 24,515 & 46 \\
\hline
S18 & riskware & Riskware& \multirow{14}{*}{Harmful Threats} &  1841 & 1353 & 14 \\
S19 & spr & SPR &  &1,789 & 1,789 &2 \\
S20 & .*deng.* & Deng & &2,926 & 2,926 & 1 \\
S21 & .*smsreg & SMSreg & & 649 & 440 & 16 \\
S22 & {[}os{]}*covav? & Cova & & 1,564 & 1,296 & 5\\
S23 & .*denofow.* & Denofow & & 1,224 & 610 & 11\\
S24 & {[}os{]}*fakeflash & FakeFlash & &  1,381 & 510 & 15 \\
S25 & .*fakeapp.* & FakeApp & & 518 & 420 & 14 \\
S26 & .*fakeinst.* & FakeInst & & 493 & 401 & 22\\
S27 & .*appinventor.* & Appinventor & & 4,025 & 3,113 & 6 \\
S28 & .*swf.* & SWF & & 4,651 & 4,566 & 10 \\
S29 & .*troj.* & Trojan (gen) & & 23,775 & 16,851 & 49 \\
S30 & .*mobi.* & Mobidash &  &  981 & 796 & 16 \\
S31 & .*spy.* & Spy && 1483 & 1,221 & 26 \\
S32 & .*gin{[}ger{]}*master & Gingermaster & & 58 & 36 & 10 \\
\hline
S33 &  unclassifiedmalware & UnclassifiedMalware & \multirow{9}{*}{Unknown/Generic} & 857 & 855 & 1 \\
S34 & .*virus.* & Virus & & 959 & 896 & 15 \\
S35 & .*heur.* & Heur & & 182 &179 & 15 \\
S36 & .*gen.* & GEN & & 9,827 & 9,118 & 25\\
S37 & {[}osgen{]}*pua & PUA & & 1,249 & 1,152 & 2 \\
S38 & {[}ws{]}*reputation & Reputation & & 2,886 & 2,885 &1 \\
S39 & .*applicunwnt.* & AppUnwanted & & 4,863& 4,860 & 1\\
S40 & .*artemi.* & Artemis & & 9,662 & 6,175 & 2 \\
S41 & .* (Default Case) & Other & & 10,778 & 7,880 & 57 \\
\hline
& TOTAL  &  & & 259,608  && \\ 
\hline
\end{tabular}
\caption{Regular Expressions in Python syntax to normalize signatures into standardized classes}
\label{tab:regexp}
\end{table*}

Table~\ref{tab:regexp} shows $41$ malware signature-based classes ($S1,\ldots,S41$) obtained using the previous methodology. The table contains the predicate of the regular expression for each rule, the class and a broader category of malware, along with the detection and application counts of each rule. 
For instance, $S1$ contains all the cases of AirPush class, which belongs to the Adware category. The AirPush class has been found in 12,802 Android apps and received 35,850 detections from 26 different AV engines.

The following lists a short summary of the three broad categories, namely emph{Adware}, \emph{Harmful Threats} and \emph{Unknown/Generic}, along with an explanation of the classes in each category.
\begin{itemize}
\item \textbf{Adware} This category includes those malware classes showing abusive advertisements for profit. There are in total $60,538$ applications tagged with at least one adware class. The \emph{Adware} category involves most apps in the collection, suggesting than most malicious applications inside Google Play are adware-related apps. \emph{Leadbolt}, \emph{Revmob}, \emph{Startapp}, \emph{WAPSX}, \emph{Dowgin/dogwin}, \emph{Cauly}, \emph{Modwin} and \emph{Apperhand/Counterclank} are well-known advertisement networks maliciously used to perform full screen and invasive advertising. \emph{Kuguo}, is an advertisement library also known due to the abuses committed by their developers. \emph{Youmi} and \emph{DroidKungFu} are advertising services which have been involved in data ex-filtration problems. \emph{Airpush} is another advertisement network company known for the abuse of its developer of adbar pushing notifications. Some AVs just mark as \emph{Multiads} applications that contain different advertisement libraries capable of displaying invasive ads. \emph{Fraud/osoneclick} refers to a fraudulent application which attempts to increase number of ad clicks in the app by stealthily settling ads in the background of user interactive applications. Finally, the \emph{Adware (gen)} tag is a generic reference assigned to those samples only containing that known class.
\item \textbf{Harmful Threats}: This category includes more dangerous threats than simple adware, which may enrol the user in premium services or ex-filtrate data through permission abuses or other exploits. There are $29,675$ applications labelled at least once in this category. \emph{Deng}, \emph{SPR (Security and Privacy Risk)} and \emph{Riskware} are generic names given by different engines to flag apps that may unjustifiably require potentially harmful permissions or include malicious code threatening user privacy. \emph{Denofow} and \emph{Cova} are generic references to trojan programs which attempts to enroll users in premium SMS services. \emph{SMSReg} is a generic way for some engines to flag applications that require SMS  related permissions for ex-filtration or premium subscription. \emph{FakeFlash}, \emph{FakeInst} or \emph{Fakeapp} are names for applications that replicate the functionalities of other popular apps adding to their code malicious code or actions. \emph{Appinventor} is a developer platform used to build and generate applications extensively preferred by malware developers. \emph{SWF} stands for different versions of Shockwave Flash Player Exploits. \emph{Trojan (gen)} is the generic reference of engines to trojan applications. \emph{GingerMaster} is a well-known family of rooting exploits. \emph{Spy} is a generic reference to applications incurring in data ex-filtration or similar spyware threats.
\item \textbf{Unknown/Generic}: This category includes AV detections which do not include class-related information, either due to generic signatures from AVs or signatures not matching any rule in the dataset. There are $23,915$ applications within this group. \emph{UnclassifiedMalware}, \emph{Virus}, \emph{Heur} (from heuristics), \emph{GEN} (Generic Malware), \emph{PUA} (Potentially Unwanted Application), \emph{Reputation}, \emph{AppUnwanted} (Application Unwanted) and \emph{Artemis} are generic tags given by different engines in order to flag applications that are detected as not-specified threats. \emph{Other} includes the remaining applications which have not been classified due to the lack of signature patterns.
\end{itemize}

As shown in the table, most common malware detection classes are typically those regarding Adware, in particular \emph{Revmob}, \emph{Airpush} and \emph{Adware} with many AVs involved. \emph{Trojan} detections are also very popular with 49 engines involved. In general, many malware classes are spotted by more than a single engine, with some exceptions in the Unknown/Category category classes, which are often exclusive from a small subset of engines, namely $S33$, $S38$, $S39$ with only one AV engine involved.

\subsection{Comparison with AVClass~\cite{avclass}}

We cloned the AVClass~\cite{avclass} repository from Github\footnote{Available at \url{https://github.com/malicialab/avclass}, last access May, 2017} and checked its performance in our dataset. We observed that the AVClass system returned undetermined class (SINGLETON output) for $48,743$ cases in our dataset, which is more than 50\% of the signatures. Oppositely, our methodology and AVClass agree on $24,097$ applications, roughly 29\% of the dataset. In this light, both approaches provide some level of agreement, as most specific classes match frequently within the clearly defined detections.





However, AVClass returns a single class per application, but does not specify which AV engine is behind such decision. In our methodology, we keep the (AV engine, Malware class) pair to allow further analysis since, in some cases, different AVs disagree on some application (some may consider it as Adware while others consider the same application as Harmful Threat for instance).


\subsection{Some insights from detections}
\label{sec:matrices}

Let  $A$ denote an indicator matrix of size $82,866\times 61$ whose elements $A_{ij}\in\{0,1\}$ are set to 1 if the $i$-th Android app has been flagged by the $j$-th engine or 0 otherwise. Matrix A is indeed very sparse with only 5\% of all the entries set to one. On average, each application is detected $3.135 \pm 3.46$ engines, showing that the variability of application detection counts is enormous. The most active AV engines are AV27, AV58, AV7, AV2, AV30 and AV32, accounting for more than $10,000$ detections each.

\begin{figure}[!htbp]
\centering{\includegraphics[scale=0.9,width=\columnwidth]{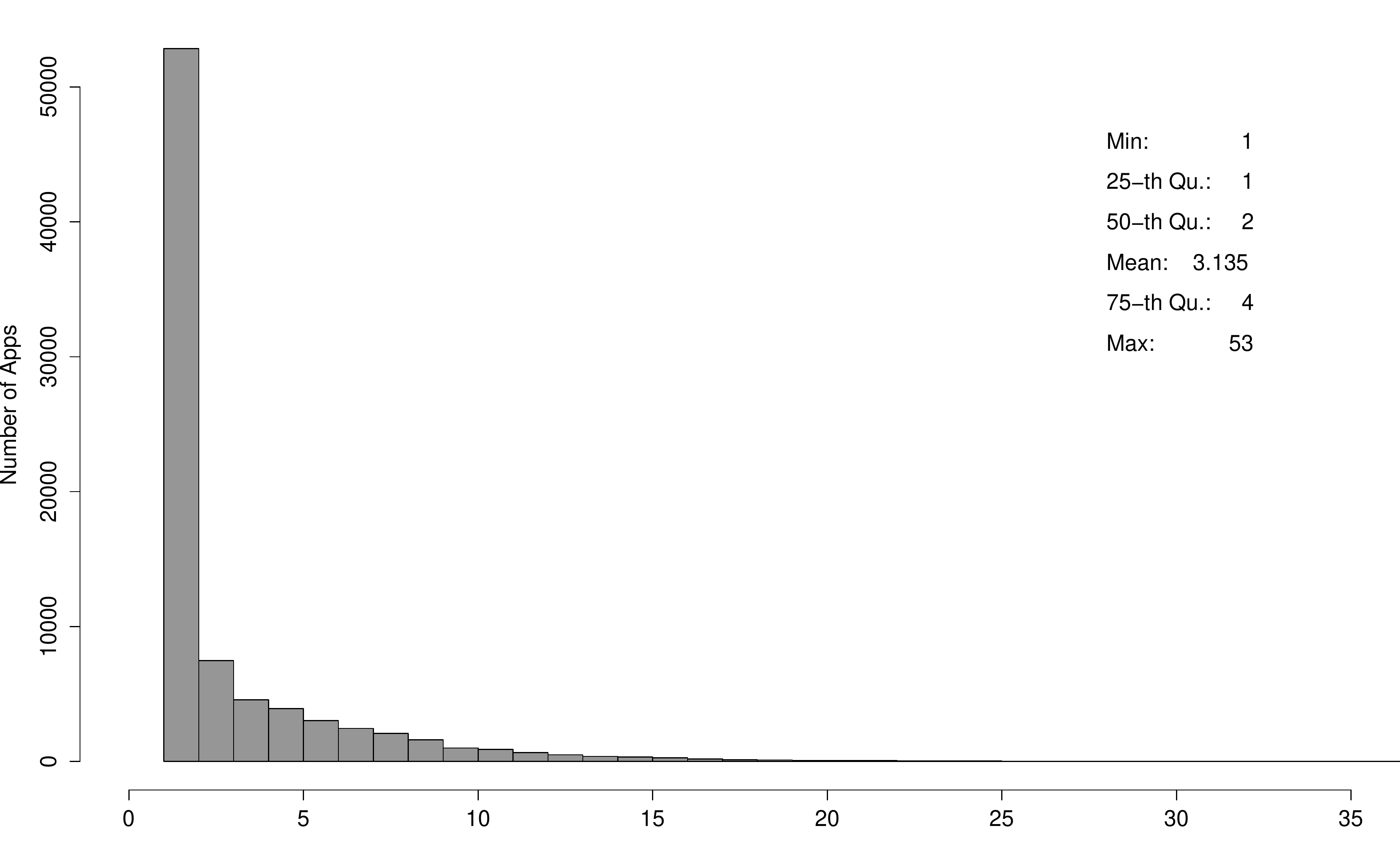}}
\caption{AV detection count per application}
\label{hist:AVflags}
\end{figure}

Fig.~\ref{hist:AVflags} is a histogram of every application detection count in matrix $A$ (a histogram of the row-sums). The histogram shows a heavy-tail like distribution where most malware applications account for a small number of detections whilst some few get much higher counts. Single-detection applications represent the majority of cases with a total of $38,933$ (46.9\% of the total). In fact, no single application is flagged by the $61$ AV engines at once, being the highest detection count for application no. $78,692$ with $53$ hits.


Now, let $B$ denote an indicator matrix of size $82,866\times 41$ whose elements $B_{ij}\in\{0,1\}$ are set to 1 if the $i$-th Android app has been flagged in the $j$-th malware category or 0 otherwise. 

\begin{figure}[!htbp]
\centering
\includegraphics[width=0.9\columnwidth]{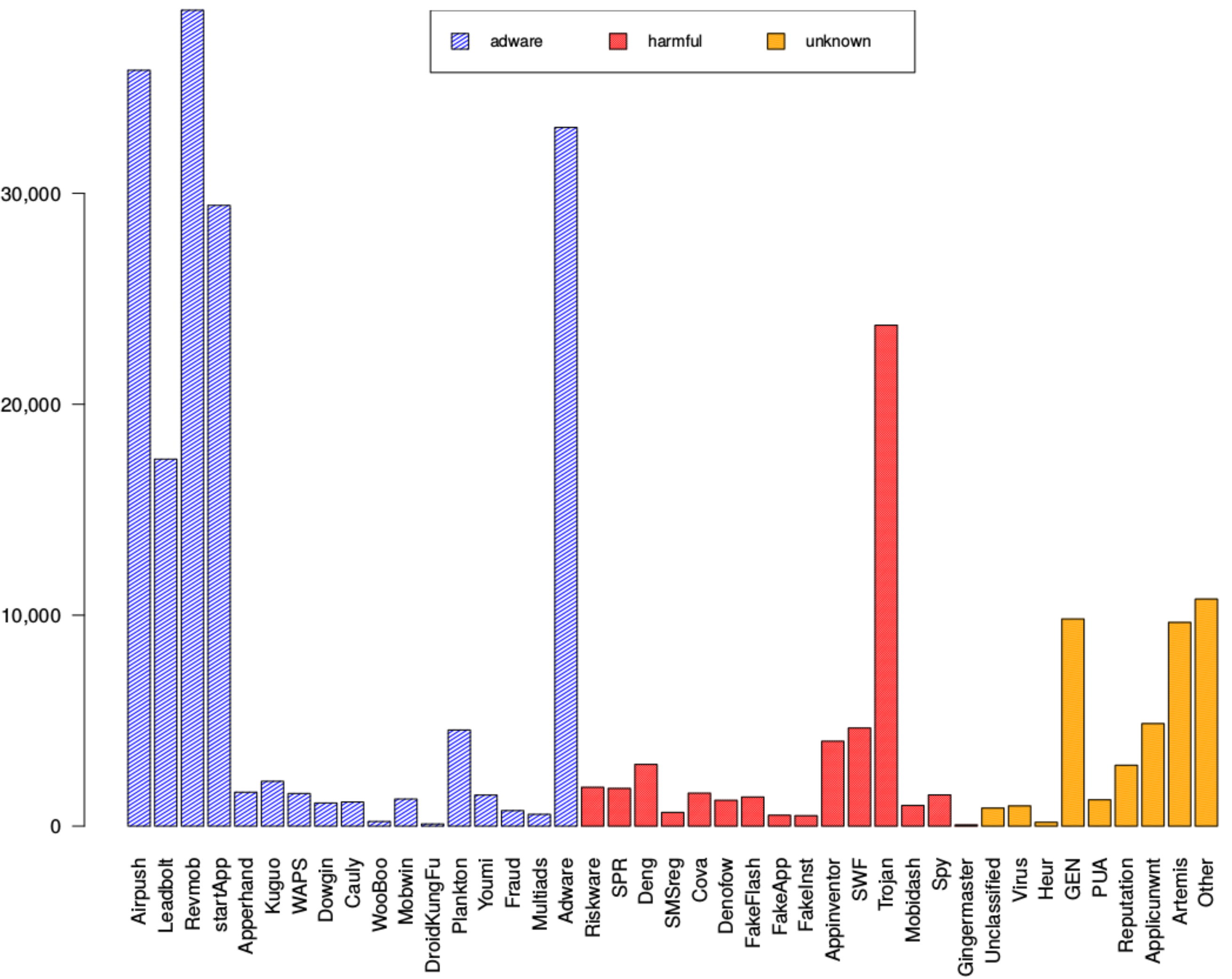}
\caption{Frequency of detections per malware class}
\label{fig:labels}
\end{figure}

Fig.~\ref{fig:labels} represents the occurrence of each malware class. At a glance, the most common classes are adware-related classes: generic adware applications and some precise libraries, namely \emph{Airpush}, \emph{Leadbolt}, \emph{Revmob} and \emph{StartApp}. The reminding detections are more infrequent, with the exception of generic \emph{Trojan} applications. Therefore, most malware applications in this collection appear to be adware cases. 

Concerning matrix $B$, we observe that out of the $43,933$ multi-detection applications, 63.26\% of them are assigned to more than one class. In particular, these $27,781$ applications receive between 2 and 12 different class labels, showing some level of disagreement between AV engines.

\section{Analysis of Malware classes and categories}
\label{sec:grouping}

As stated before, there are $38,933$ Android apps flagged by a single AV engine. Of the rest (those with two AV detections or more), in $16,152$ cases all AV detections agree on the same malware class, while the remiainng $27,781$ apps show some kind of disagreement between AVs. In fact, some authors have proven the existing lack of consensus of engines~\cite{lackConsensus} as well as severe class naming inconsistencies among engines~\cite{nameInconsistencies}. 

In this section, we analyze whether any of the inferred classes differ just because of naming inconsistencies or they represent a set of independent classes. 


\subsection{Correlation of malware categories}

Remark that the 41 malware classes have been classified into three large malware categories, namely \emph{Adware}, \emph{Harmful Threats} and \emph{Unknown/Generic}. The three categories are very broad and the nature of the malware involved can be different. Nevertheless, the detections in both Adware and Harmful categories separate malware into low-risk and high-risk malware classes, as Adware samples are typically controversial and not detected by all the engines in the same way whilst harmful classes indicate potentially major security risks, such as data leakage or economic loss.

In addition, the Unknown/General category integrates all those malware classes which do not refer to any specific malware type, being just an indicator of undesired behaviors. 

Let  $D$ refer to an $82,866\times 3$ matrix where $D_{ij}$ is an integer which accounts for the number of times the $i$-th application has received a detection in category Adware ($j=1$), Harmful ($j=2$) or Unknown ($j=3$). Table~\ref{tab:catcorr} shows the correlation of such matrix $D$.


\begin{table}[!htbp]
\begin{tabular}{|c| c c c |}
\hline
&Adware&Harmful&Unknown\\
\hline
Adware&1&0.06&0.3\\
Harmful&0.06&1&0.44\\
Unknown&0.3&0.44&1\\
\hline
\end{tabular}
\caption{Correlation of matrix $D$ (Malware Categories)}
\label{tab:catcorr}
\end{table}

As shown, Harmful and Adware categories show little correlation, only 0.06 which may refer to  Android apps both presenting Adware and being potentially Harmful. On the other hand, Unknown/Generic apps show 0.3 correlation with Adware and 0.44 correlation with Harmful Threats. 

Interestingly, it seems that Unknown detections flagged by some AV engines appear more often with Harmful Threats by other AV engines than with Adware cases, showing that Unknown detections are probably cases of Harmful Threats. This shall be further investigated in Section~\ref{sec:logReg}.

\subsection{Identifying relationships with classes with graph community algorithms}


Graph theory provides useful algorithms to study the relationships between objects within a network of entities. In our case, starting from matrix $B$ defined in Section~\ref{sec:matrices} we compute its correlation matrix, i.e. $Corr(B)$ and define a Graph $G=(N,E)$ whose adjacency matrix is $Corr(B)$. Thus, graph G has 41 nodes (malware classes) and the weights of the edges are equal to the correlation values between malware classes.

Using node edge betweenness~\cite{edgeBetweeness}, we group together nodes according to their correlation values to see which malware classes are close together. In order to avoid generating communities out of noise, we force all correlation values below some $Corr_{min}$ threshold to be equal to 0. 


\begin{figure*}[!ht]
\centering
\subfigure[$Corr_{min}=0.2$]{
\includegraphics[width=0.45\textwidth]{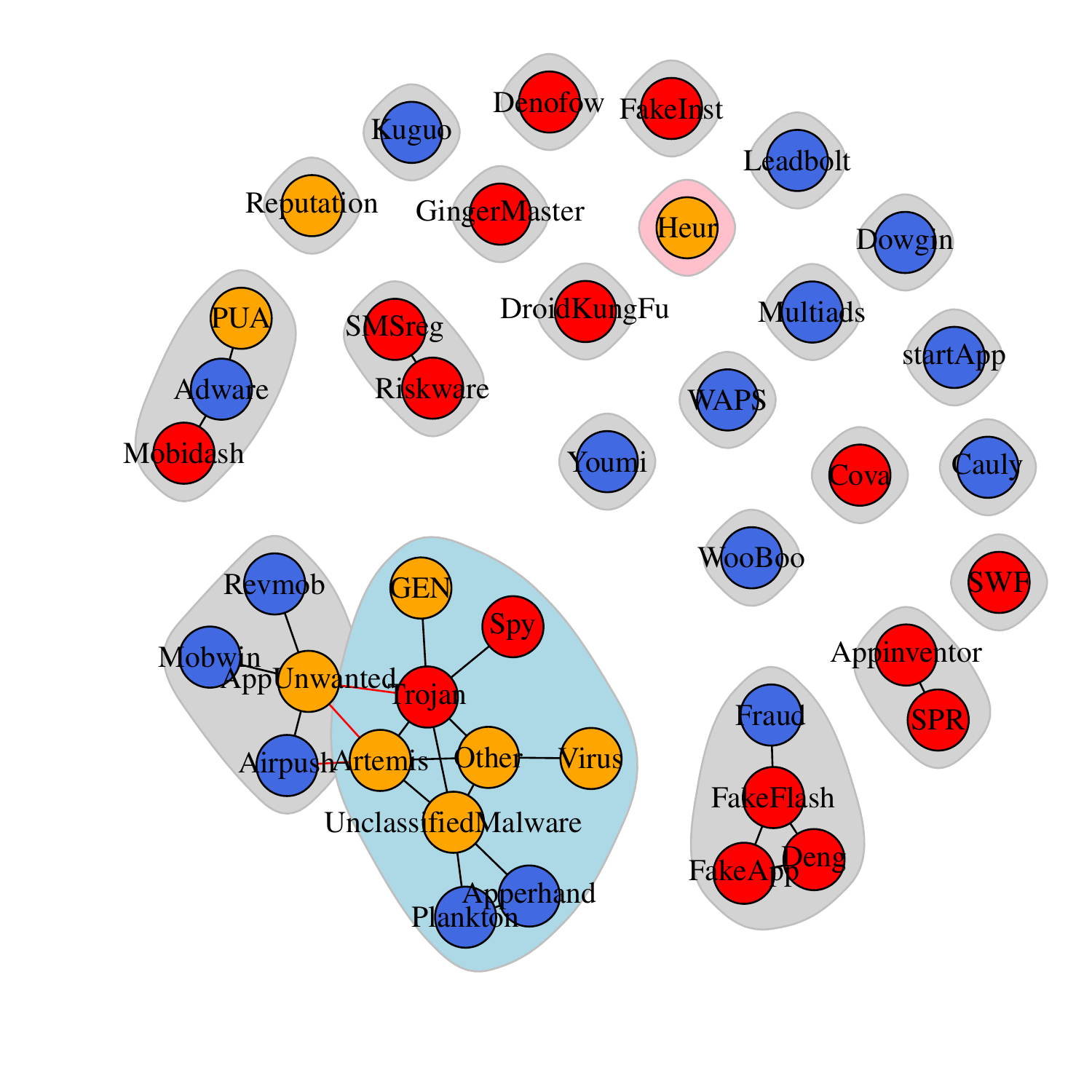}
}
\subfigure[$Corr_{min}=0.35$]{
\includegraphics[width=0.45\textwidth]{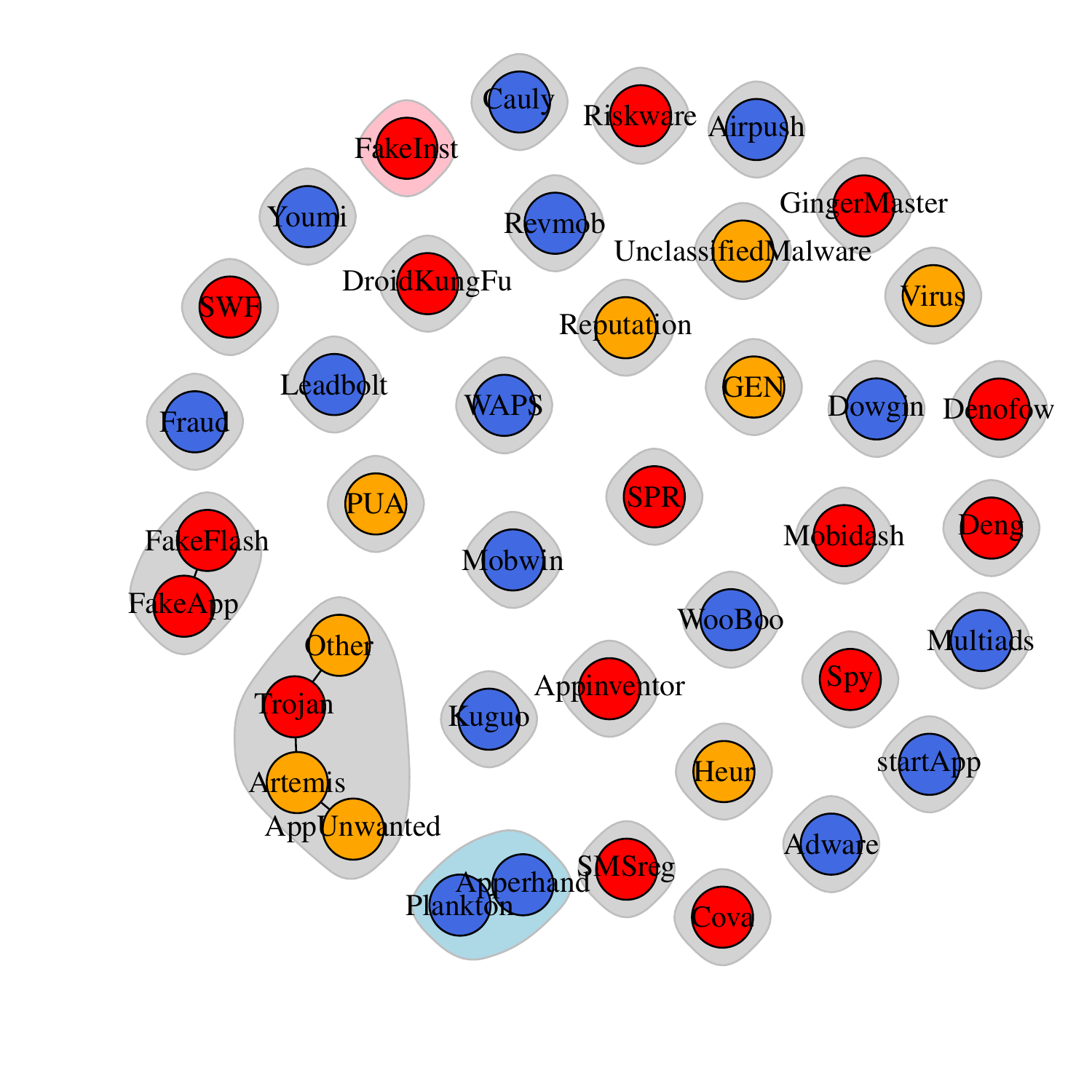}
}
\subfigure[$Corr_{min}=0.5$]{
\includegraphics[width=0.45\textwidth]{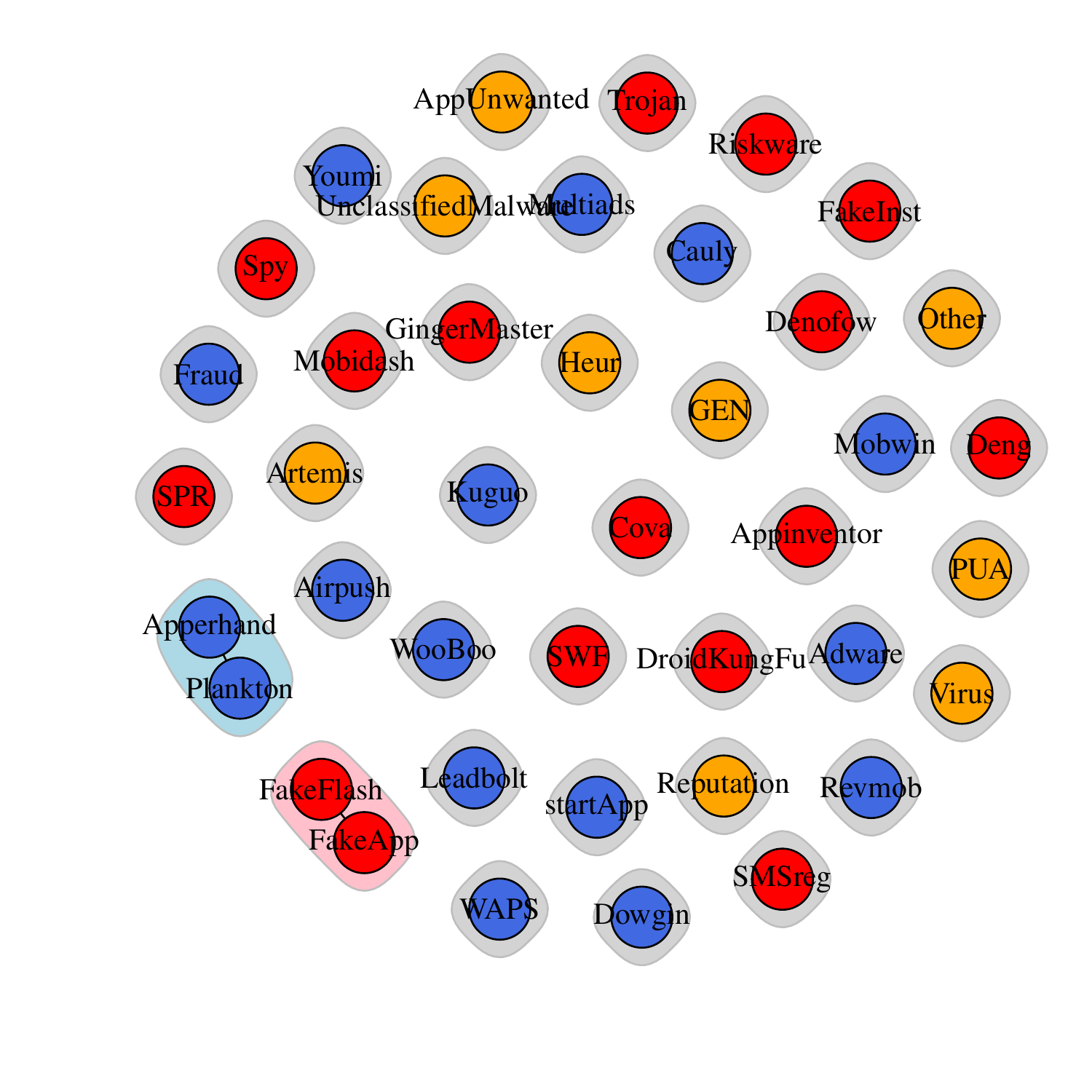}
}
\caption{Communities of malware classes for different correlation threshold $Corr_{min}$}
\label{fig:graph_sig}
\end{figure*}

Fig.~\ref{fig:graph_sig} depicts the resulting communities obtained. Essentially, most malware classes appear isolated with little relationships with others, especially when the correlation threshold $Corr_{min}=0.5$. In such a case, only two new communities are created: \emph{FakeFlash-FakeApp} and \emph{Plankton-Apperhand}, the former in the Harmful category, the latter in the Adware category. In case of a lower correlation threshold, $Corr_{min}=0.35$ a new community is identified: three classes belonging to the Unknown/Generic category are aggregated with a Harmful class, creating the community \emph{Trojan-Artemis-AppUnwanted-Other}. This is consistent with the previous experiment where we observed that most Unknown cases are more correlated with Harmful than with Adware cases. Finally, when the correlation threshold $Corr_{min}=0.2$ we observe other interesting communities with weak correlation.

\section{Modelling Consensus}
\label{sec:logReg}


In this section, we further investigate on AV engines and malware categories using Structure Equation Models (SEM) to identify which AVs are more powerful at detecting Adware, Harmful Threats and Unknown/Generic categories.

%
%
%
%

\subsection{On weighting AV engines}

In order to obtain a performance score per engine within our dataset, a collective AV model must consider how AV engines behave collectively in the sense that which AV engines are consistent with other and which ones typically disagrees with the rest. This idea is at the heart of the well-known Latent Variable Models (LVM) which assume the existence of some unobservable "latent" or "hidden" variable (i.e. whether an app is of malware category or not) which explains the underlying relation among the observed variables (i.e. the output of the AV engines). 

There exist different approaches to Latent Variable Modeling in the literature, such as generative models or Structural Equation Models (SEM). We have chosen the later due to its ease of use and approach, based on covariance approximation, which weight engines according on how consensual their detections are.
 
Typically, SEM assumes a linear regression model on the latent or hidden variable, namely $Z_{sem}$:
\begin{equation}
Z_{sem}=\sum_{i=1}^{61} \omega_{i}\times X_{AVi}
\label{eq:model}
\end{equation}
where $X_{AVi}$ refers to the observed variables AVi weighted by coefficients $\omega_i$. In order to shape values to a probabilistic scale, we use the logistic function to translate the $Z_{sem}$ score into a probabilistic value (between 0 and 1), following:
\begin{equation}
P_{sem} = \frac{e^{Z_{sem}}}{1+e^{Z_{sem}}}
\label{eq:logistic}
\end{equation}



\subsection{Inference and Results}

\begin{figure*}[!htbp]
\centering
\includegraphics[width=\textwidth]{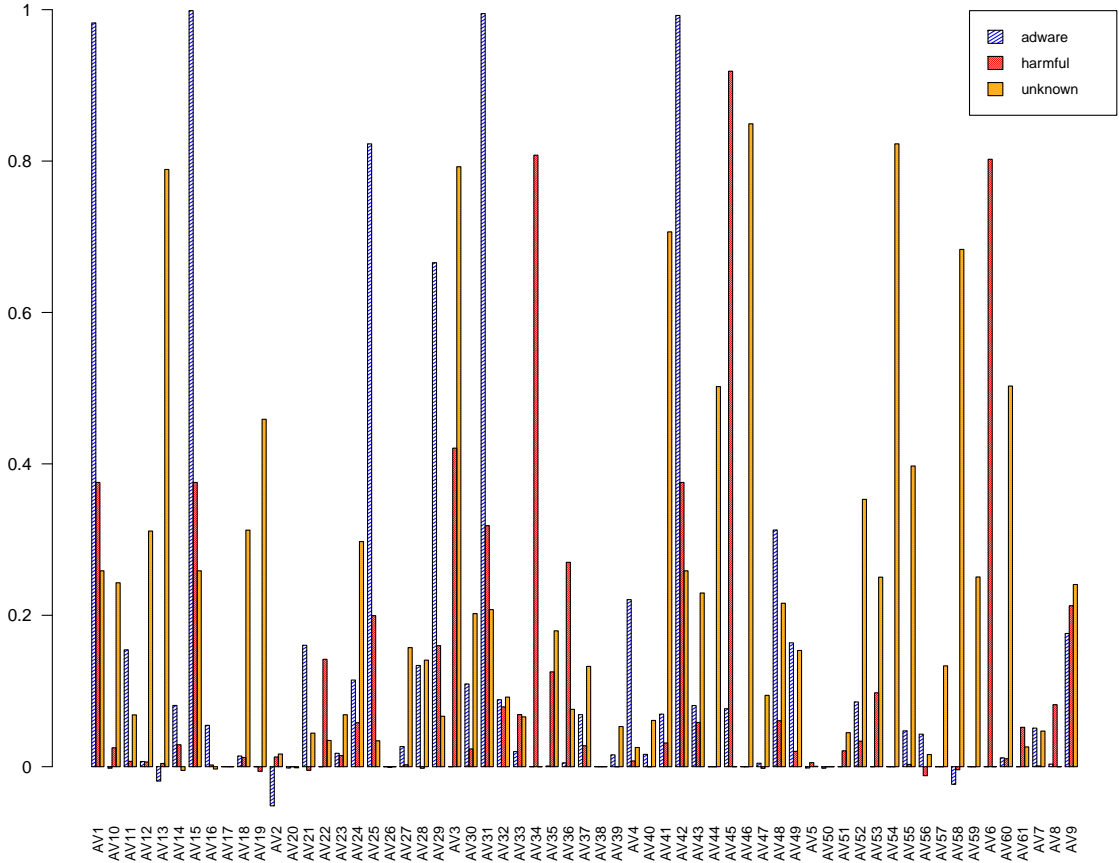}
\caption{$Z_{SEM}$ coefficients for each category}
\label{fig:results}
\end{figure*}

We generate three 0/1 matrices, one per category (Adware, Harmful Threats and Unknown/Generic), of size $Apps \times AVs$. These matrices are used for training three $Z_{sem}$ models using the R-library "lavaan"~\cite{lavaan}, which estimates the $\omega$ coefficients by minimizing the difference between the dataset covariance matrix and the covariance matrix from the generated model. The $\omega$ coefficients are shown in Fig.~\ref{fig:results} for the three models.

The figure clearly unveils the existing differences across engines: some AVs rank high scores at specific categories, while others are terrible on all three categories. For instance, AV6 excels at harmful applications (coefficient 0.8) but has null properties (coefficient 0) for Adware or Unknown malware; AV6 is very good at Harmful Threats (coefficients 0.8 for harmful threats and 0 for the other two categories) and AV41 is excellent with unknown categories (0.7). Other AV engines have acceptable coefficients for more than one category, such as AV1 or AV15.  In fact, adware-detecting engines, appear with very high coefficients  whereas unknown detections occur notably across most engines.

The picture also shows clearly that there is no AV engine in this collection which excels in the three categories at the same time, even though, engines such as AV31 (0.99, 0.31, 0.2) or AV42 (0.99, 0.37, 0.25) show some of the best balances above all engines. Indeed, AV1, AV15, AV31 and AV42 present strong agreements and strong correlation values, providing high support from one another.

It is also worth noting the spikes of some AV engines at the Unknown category (see AV13, AV46, AV54, AV58 or AV60). Essentially, these engines do not provide information on Adware or Harmful Threats, instead they output a generic malware signature, and therefore receive low weights on such categories. 


Finally, as an example of application of the $Z_{sem}$ models, consider app no. $1,144$. This app has been flagged as Adware by 20 AVs, as Harmful by AV47, and Unknown by AV22, AV39 and AV40. We can then apply eq.~\ref{eq:model} to obtain the $Z_{sem}$ values for the three categories which yields:

\begin{eqnarray}
\nonumber Z_{sem}^{(1,144,Adware)} & = &  4.66 \\
\nonumber Z_{sem}^{(1,144,Harmful)} & = &  3.8\\
\nonumber Z_{sem}^{(1,144,Unknown)} & = & 3.91\\
\end{eqnarray}

After applying the logistic transformation of eq.~\ref{eq:logistic}, we obtain the following probabilities: 
\begin{eqnarray}
\nonumber P_{sem}^{(1,144,Adware)} & = & 0.99 \\
\nonumber P_{sem}^{(1,144,Harmful)} & = & 0.97 \\
\nonumber P_{sem}^{(1,144,Unknown)} & = & 0.98 \\
\end{eqnarray}

In this case, it would be safe to say that this application lies in all categories of malware: Adware, Harmful and Generic. 

On the other hand, Android app no. $11,581$ has been flagged as Adware by AV7, AV14 and AV36, as Harmful by AV4, and Unknown by AV36. According to the SEM model, the following probabilities for each category are obtained:
\begin{eqnarray}
\nonumber P_{sem}^{(11,581,Adware)} & = & 0.588 \\
\nonumber P_{sem}^{(11,581,Harmful)} & = & 0.61 \\
\nonumber P_{sem}^{(11,581,Unknown)} & = & 0.54 \\
\end{eqnarray}
Checking the coefficients of these AV engines on each category in Fig.~\ref{fig:results}, we observe that AV7, AV14 and AV36 have low coefficients for Adware, while AV4 and AV36 also have low coefficient values for Harmful and Unknown respectively. Perhaps, the Harmful category is more likely according to the estimation provided by the SEM mmodel despite the app has more Adware detections. However, it is not clear whether or not this application represents a real risk to the user. 

As a final example, Android app $67,119$, which accounts 14 detections clearly votes on favor for the Adware category:
\begin{eqnarray}
\nonumber P_{sem}^{(67,119,Adware)} & = & 0.91 \\
\nonumber P_{sem}^{(67,119,Harmful)} & = & 0.62 \\
\nonumber P_{sem}^{(67,119,Unknown)} & = & 0.89 \\
\end{eqnarray}

\section{Summary and conclusions}
\label{sec:conclusions}

This paper has analyzed $259,608$ malware signatures produced by 61 AV engines to $82,866$ different Android applications. With this dataset, we have:
\begin{itemize}
\item Presented a novel signature normalization methodology capable of mapping different AV signatures into $41$ standardized \emph{classes}.
\item Analyzed the most frequent keywords and signature categories using text mining and minhashing techniques, and classified malware signatures into three categories: Adware, Harmful threats and Unknown/Generic.
\item Identified groups of similar malware classes within the data using Community detection algorithms from Graph Theory. 
\item Used Structural Equation Models to find most powerful AV engines for each of the three malware category.
\item Shown an application on how to use such SEM model to infer which Unknown-type applications are closer to Adware or Harmful type.
\end{itemize}

\section*{Acknowledgments}
The authors would like to acknowledge the support of the national project TEXEO (TEC2016-80339-R), funded by the Ministerio de  Econom\'{i}a y Competitividad of SPAIN through, and the EU-funded H2020 TYPES project (grant no. H2020-653449).

Similarly, the authors would like to remark the support provided by the tacyt system (\url{https://www.elevenpaths.com/es/tecnologia/tacyt/index.html}) for the collection and labelling of AV information.

\bibliographystyle{ACM-Reference-Format}
\bibliography{all}
\end{document}